\documentclass[american,aps,prl,reprint,longbibliography,superscriptaddress]{revtex4-1}

\usepackage{lineno}
\usepackage{subfigure}
\usepackage{sublabel}
\usepackage{dcolumn}
\usepackage{amsmath,amssymb}
\usepackage{enumitem}

\usepackage{bm}
\usepackage{bbm}
\usepackage{color}
\usepackage{overpic}
\usepackage{latexsym}
\usepackage{epstopdf}
\usepackage{color}
\usepackage[english]{babel}
\usepackage[utf8]{inputenc}
\usepackage{times}
\usepackage{latexsym}
\usepackage{stmaryrd}
\usepackage{psfrag,graphicx}
\usepackage{epsf}
\usepackage{subfigure}
\usepackage{amsmath}
\usepackage{amssymb}
\usepackage{amsfonts}
\usepackage{natbib}
\usepackage{epstopdf}
\usepackage{appendix}
\usepackage{enumitem}
\usepackage{braket} 				%Dirac-Notation

\newcommand{\ketbra}[2]{\vert#1\rangle\!\langle#2\vert}

\def\ket#1{\left|#1\right\rangle}

\definecolor{mygrey}{gray}{0.35}
\definecolor{myblue}{rgb}{0.2,0.2,0.8}
\definecolor{myzard}{cmyk}{0,0,0.05,0}
\definecolor{mywhite}{rgb}{1,1,1}
\definecolor{myred}{rgb}{0.9,0.1,0.}
\usepackage[colorlinks=true,citecolor=myblue,linkcolor=myblue,urlcolor=myblue]{hyperref}

\begin{document}
\title{Efficient Distinction between Quantum Direct and Common Causes\\ and its Experimental Verification}

\author{Feixiang Xu}
\thanks{These two authors contributed equally}
\affiliation{National Laboratory of Solid State Microstructures,
College of Engineering and Applied Sciences and School of Physics, Nanjing University, Nanjing 210093, China}
\affiliation{Collaborative Innovation Center of Advanced Microstructures, Nanjing University, Nanjing 210093, China}

\author{Jia-Yi Lin}
\thanks{These two authors contributed equally}
\affiliation{National Laboratory of Solid State Microstructures,
College of Engineering and Applied Sciences and School of Physics, Nanjing University, Nanjing 210093, China}
\affiliation{Collaborative Innovation Center of Advanced Microstructures, Nanjing University, Nanjing 210093, China}
\affiliation{Institute for Brain Sciences and Kuang Yaming Honors School, Nanjing University, Nanjing 210023, China}

\author{Ben Wang}
\affiliation{National Laboratory of Solid State Microstructures,
College of Engineering and Applied Sciences and School of Physics, Nanjing University, Nanjing 210093, China}
\affiliation{Collaborative Innovation Center of Advanced Microstructures, Nanjing University, Nanjing 210093, China}

\author{Tao Jiang}
\affiliation{National Laboratory of Solid State Microstructures,
College of Engineering and Applied Sciences and School of Physics, Nanjing University, Nanjing 210093, China}
\affiliation{Collaborative Innovation Center of Advanced Microstructures, Nanjing University, Nanjing 210093, China}

\author{Shengjun Wu}
\email{sjwu@nju.edu.cn}
\affiliation{National Laboratory of Solid State Microstructures,
College of Engineering and Applied Sciences and School of Physics, Nanjing University, Nanjing 210093, China}
\affiliation{Collaborative Innovation Center of Advanced Microstructures, Nanjing University, Nanjing 210093, China}
\affiliation{Institute for Brain Sciences and Kuang Yaming Honors School, Nanjing University, Nanjing 210023, China}

\author{Wei Wang}
\email{wangwei@nju.edu.cn}
\affiliation{National Laboratory of Solid State Microstructures,
College of Engineering and Applied Sciences and School of Physics, Nanjing University, Nanjing 210093, China}
\affiliation{Collaborative Innovation Center of Advanced Microstructures, Nanjing University, Nanjing 210093, China}
\affiliation{Institute for Brain Sciences and Kuang Yaming Honors School, Nanjing University, Nanjing 210023, China}

\author{Lijian Zhang}
\email{lijian.zhang@nju.edu.cn}
\affiliation{National Laboratory of Solid State Microstructures,
College of Engineering and Applied Sciences and School of Physics, Nanjing University, Nanjing 210093, China}
\affiliation{Collaborative Innovation Center of Advanced Microstructures, Nanjing University, Nanjing 210093, China}

\begin{abstract}
Identifying the causal structures between two statistically correlated events has been widely investigated in many fields of science. While some of the well-studied classical methods are carefully generalized to quantum version of causal inference for certain cases, an effective and efficient way to detect the more general quantum causal structures is still lacking. Here, we introduce a quantity
named `Causal Determinant' to efficiently identify the quantum causal structures between two quantum systems and experimentally verify the validity of the method. According to the causal determinant, the quantum direct cause imposed by an arbitrary unitary operator can be perfectly discriminated with the quantum common cause, in which the two quantum systems share a joint quantum state. In addition, the causal determinant has the capability to discriminate between more general causal structures and predict the range of their parameters. The ability to detect more general quantum causal structures of our method can shed new light on the field of quantum causal inference.
\end{abstract}
\maketitle

\begin{figure}[t]
	 \centering
   \includegraphics[width=0.4\textwidth]{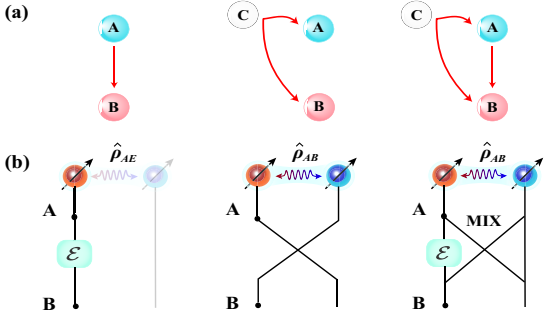}
   \caption[width=1\textwidth]{(a) The directed acyclic graphs (DAGs) for quantum direct causes, quantum common causes and mixtures of them (from left to right). (b) The quantum circuits corresponding to the three causal structures. The direct cause is described by a local evolution $\mathcal{E}$. The possible ancillary system $E$ that couples to $A$ is not involved in the causal relation between $A$ and $B$. The common cause is represented by the shared bipartite state $\hat{\rho}_{AB}$. $\mbox{MIX}$ stands for the probabilistic mixture of the two causal mechanisms.}
	\label{fig:schematic}
\end{figure}

Due to the fact that `correlation does not imply causation' \cite{reichenbach1991direction}, determining the causal structure between correlated events is a fundamental but very hard problem \cite{pearl2009causality,spirtes2000causation}. Even for the simplest condition shown by the directed acyclic graphs (DAGs) in Fig. \ref{fig:schematic}(a), where the causal structures between the two correlated events $A$ and $B$ can be: $A$ is the direct cause of $B$, $A$ and $B$ share a common cause $C$ and the mixture of the two, the completely identification is still complicated. A typical example is that in medical science the doctor tends to infer the causality (direct cause) between the drug and the recovery of a disease in the presence of certain common causes, e.g., the gender of the patients which may affect both the inclination to take the treatment and the recovery. Several classical methods, like interventions and instrumental tests, are developed to distinguish the effects of the direct cause and common cause \cite{peters2017elements,shipley2016cause,morgan2015counterfactuals}. However, these methods cannot be applied to the quantum version of causal inference straightforwardly due to the peculiar features of the quantum mechanics \cite{PhysRevLett.114.060405}, i.e., the violation of the Bell inequality \cite{PhysicsPhysiqueFizika.1.195,PhysRevLett.106.200402,PhysRevA.88.052127}.
The instrumental inequalities, which are introduced to estimate classical causal influence without using intervention \cite{balke1997bounds}, are violated by the quantum correlations \cite{chaves2018quantum,van2019quantum,PhysRevLett.120.140408,PhysRevLett.125.230401}. These results indicate that quantum correlations do not comply to the classical causal constraints. Therefore, the definitions of and methods to describe the classical causal structure are carefully generalized to the quantum realm \cite{tucci1995quantum,tucci2007factorization,PhysRevA.74.042310,LEIFER20081899,PhysRevA.88.052130,Henson_2014}, motivating the development of quantum causal models \cite{oreshkov2012quantum,brukner2014quantum,Pienaar_2015,procopio2015experimental,chaves2015information,Costa_2016,PhysRevX.7.031021,rubino2017experimental,pienaar2019time,barrett2019quantum,giarmatzi2018quantum,chiribella2019quantum,lee2017causal,maclean2017quantum,PhysRevA.95.062102,wolfe2019inflation,kela2019semidefinite,PhysRevLett.125.110505,barrett2020cyclic,PhysRevA.101.012104} which find applications in many quantum information tasks including quantum computation \cite{PhysRevLett.113.250402,PhysRevA.96.052315}, quantum metrology \cite{PhysRevLett.124.190503} and quantum communication \cite{PhysRevLett.117.100502,PhysRevA.92.052326,PhysRevA.86.040301,bavaresco2020strict}.

Within this quantum causal framework, recently, the quantum correlations are proved to be helpful to witness the causal structures between two quantum systems \cite{fitzsimons2015quantum,ried2015quantum}.
These methods are further developed to discriminate between the quantum direct cause and the quantum common cause for some specific cases \cite{PhysRevA.97.062125, PhysRevA.101.062103}. The situation that quantum common and direct causes contribute to the correlation simultaneously is also investigated \cite{K_bler_2018}. However, there is still no efficient way to distinguish the quantum common cause described by an arbitrary bipartite state from the quantum direct cause imposed by a unitary operator, let alone the more general quantum channels. Moreover, for the general quantum causal structures with both common and direct causes involved, a method that can evaluate their contributions remains to be explored.

In this work, we introduce a quantity
named the causal determinant, which can be efficiently calculated from the correlations between the measurement outcomes on both quantum systems. We show that any quantum direct cause imposed by a unitary operator can be perfectly discriminated with quantum common cause described by an arbitrary joint quantum state according to the values of the causal determinant. Even if the single unitary operator in the quantum direct cause is replaced by a more general quantum channel, the causal determinant can still have certain capability to distinguish the two causal mechanisms. For generalized causal structures where the quantum common cause is probabilistically mixed with the quantum direct cause, we can use the causal determinant to witness the existence of both causal mechanisms and can even determine the range of the mixing probability. A quantum-optical experiment is implemented and the results agree well with the theoretical predictions, confirming the validity of the method in characterizing quantum causal structures.

{\em Causal inference--}The causal structures considered in this work between the two quantum systems $A$ and $B$ are described by the quantum circuits shown in Fig. \ref{fig:schematic}(b). The quantum system $A$ is considered as a direct cause of the quantum system $B$ if the quantum state $\hat{\rho}_B$ is evolved from the quantum state $\hat{\rho}_A$, i.e., $\hat{\rho}_B=\mathcal{E}(\hat{\rho}_A)$.
For simplicity, we start our analysis with the situation that the evolution is unitary, i.e.,
\begin{equation}
	\hat{\rho}_B=\hat{U}\hat{\rho}_A\hat{U}^\dagger. \label{eq:causality}
\end{equation}
The quantum common cause is described by a bipartite state $\hat{\rho}_{AB}$ shared by the two systems. A more general situation may mix both causal mechanisms in a probabilistic way.

We focus on the situation that both $A$ and $B$ are qubits. To identify the causal structure, two observables $\hat{O}_A$ and $\hat{O}_B$ are measured on each system respectively, with each observable one of the Pauli operators $\hat{\sigma}_i$ ($i = 1, 2, 3$). The measurement outcomes are binary with $o_A = \pm 1$ and $o_B = \pm 1$.
When the same observables are measured on both systems, the correlations are defined as $c_{ii}=p_{ii}(o_A=o_B)-p_{ii}(o_A\neq o_B)$.
These correlations provide certain capability to discriminate between some quantum direct causes and quantum common causes \cite{ried2015quantum, PhysRevA.97.062125, PhysRevA.101.062103}. However, they are not sufficient to distinguish all possible quantum direct causes described by Eq.~(\ref{eq:causality}) without adjusting the measurement basis according to the details of $\hat{U}$~\cite{PhysRevA.101.062103}, let alone more general channels, from the quantum common cause.

For the direct cause given in Eq.~(\ref{eq:causality}), the correlations defined above coincide exactly with the diagonal elements of the inverse rotation matrix $R_{\hat{U}}^{-1}$ in the Bloch sphere representation \cite{Supplementary_material}. This coincidence motivates us to extend the correlations into nine $\{c_{jk}\}_{j,k=1}^3$ to take into account the situations that different observables $\hat{\sigma}_j$ and $\hat{\sigma}_k$ are measured on the two systems, respectively. Then we obtain the correlation matrix %$C=[c_{11}~c_{12}~c_{13};c_{21}~c_{22}~c_{23};c_{31}~c_{32}~c_{33}]$.
\begin{equation}
  C=
  \begin{bmatrix}
    c_{11}&c_{12}&c_{13}\\c_{21}&c_{22}&c_{23}\\c_{31}&c_{32}&c_{33}
  \end{bmatrix}.\label{eq:cmatrix}
\end{equation}
In the scenario of the quantum direct cause connected by a unitary operator $\hat{U}$, the correlation matrix obtained above in Eq. \eqref{eq:cmatrix}
equals to the inverse of the rotation matrix $R_{\hat{U}}^{-1}$ in the Bloch sphere representation, i.e., $C^{DC}(\hat{U}) = R_{\hat{U}}^{-1}$~\cite{Supplementary_material}. Since $R_{\hat{U}}^{-1}$ belongs to $\mathrm{SO}(3)$, the determinant of the correlation matrix should be a constant $1$, i.e., $\Delta^{DC}(\hat{U})=\text{det}~C^{DC}(\hat{U})\equiv 1$.
Here the superscript $DC$ means `direct cause' and $\Delta$ denotes the determinant of the correlation matrix, which is termed as the `Causal Determinant'.

If the two systems share common cause described by a density matrix $\hat{\rho}_{AB}$, which can be represented by two vectors $\vec v_A,~\vec v_B\in\mathbb{R}^3$ and a matrix $M\in\mathbb{R}^{3\times 3}$ with elements $\{m_{jk}\}_{j,k=1}^3$, i.e.,
$\hat{\rho}_{AB}=\frac{1}{4}[\hat{I}_A\otimes\hat{I}_B+(\vec v_A\cdot\vec{\sigma})\otimes\hat{I}_B+\hat{I}_A\otimes(\vec v_B\cdot\vec{\sigma})+\sum^3_{j,k=1}m_{jk}\left(\hat{\sigma}_j\otimes\hat{\sigma}_k\right)]$,
where $\hat{I}$ is the identity operator and $\vec{\sigma}$ is the Pauli vector $[\hat{\sigma}_1,\hat{\sigma}_2,\hat{\sigma}_3]$, the correlation matrix in Eq. \eqref{eq:cmatrix}
equals exactly to the matrix $M$ in the generalized Bloch sphere representation \cite{Supplementary_material}, i.e., $C^{CC}(\hat{\rho}_{AB})=M$. By choosing appropriate local basis, the matrix $M$ can be diagonalized so that the range of its causal determinant can be calculated as $\Delta^{CC}(\hat{\rho}_{AB})\in[-1,1/27]$. Here, we use the superscript $CC$ to denote `common cause'.

Apparently, the ranges of the causal determinant for above two causal mechanisms do not overlap with each other, i.e., $\Delta^{DC}(\hat{U})\equiv 1$ and $\Delta^{CC}(\hat{\rho}_{AB})\in[-1,1/27]$. Therefore, the direct cause given in Eq. \eqref{eq:causality} can be completely discriminated with any common cause. Using the causal determinant, this discrimination scheme is efficient without resorting to the method of the causal tomography. Furthermore, since the determinant is invariant under local unitary transformations, its value is independent on the reference frames of the local measurements on systems $A$ and $B$, which, therefore, do not have to be aligned. In addition, we generalize the extremal cases studied in Ref. \cite{ried2015quantum}. In those cases, only the three diagonal elements of correlation matrix in Eq. \eqref{eq:cmatrix}
are non-zero, whose product equals to the determinant of the correlation matrix. (See details in Ref. \cite{Supplementary_material})

{\em Generalization--}We now use the causal determinant to analyze the quantum direct cause imposed by a more general quantum channel $\mathcal{E}$, which is the convex combination of $N^{DC}$ unitary operators, i.e., $\hat{\rho}_B=\mathcal{E}(\hat{\rho}_A)=\sum_{m=1}^{N^{DC}}a_m\hat{U}_m\hat{\rho}_A\hat{U}_m^\dagger $
with $\sum_{m=1}^{N^{DC}}a_m=1$ and $a_m\ge 0$. For qubit, this quantum channel leaves the maximally mixed state invariant, which is equivalent to the unital quantum channel \cite{nielsen_chuang_2010}. In this case, the correlation matrix is $C^{DC}(\mathcal{E})=\sum_{m=1}^{N^{DC}}a_mC^{DC}(\hat{U}_m)$. The range of its causal determinant is related to the number of the unitary operators $N^{DC}$.
When this more general quantum direct cause is probabilistically mixed with the quantum common cause with the mixing probability $p$, the correlation matrix $C$ is the probabilistic mixture of each corresponding correlation matrix, i.e., $C=pC^{DC}(\mathcal{E})+(1-p)C^{CC}(\hat{\rho}_{AB})$.
In this way, the range of the causal determinant for fixed mixing probability $p$ and the number of unitary operators $N^{DC}$ can be calculated theoretically. These results can in reverse help us to infer the causal structures between the two quantum systems given an experimentally measured causal determinant.

\begin{figure*}[t]
	 \centering
   \includegraphics[width=1\textwidth]{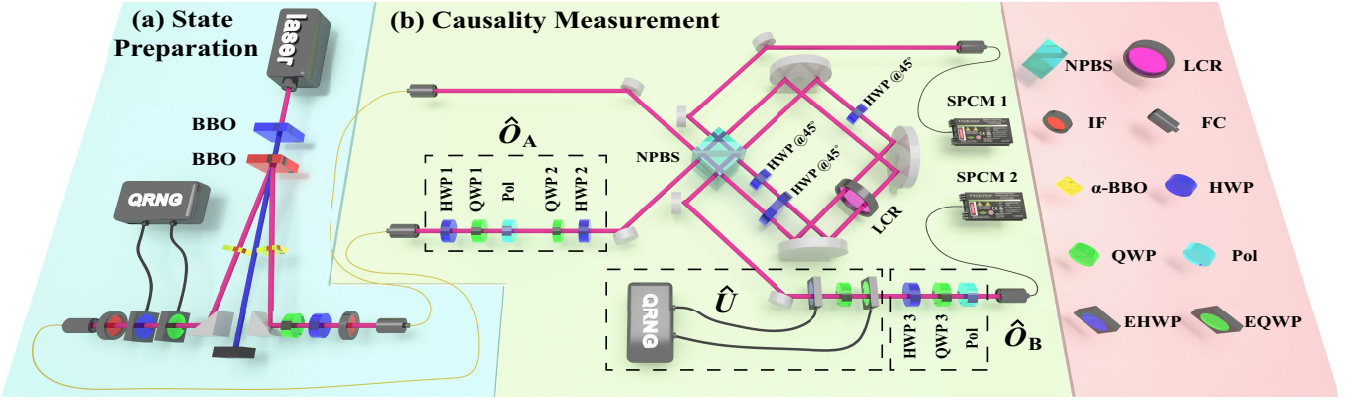}
   \caption[width=1\textwidth]{Experimental set-up. (a) State preparation. Different states are generated via spontaneous parametrical down-conversion (SPDC) with the help of a quantum random number generator (QRNG). (b) Causality measurement. The appropriate phase difference inserted between the two displaced paths in the displaced Sagnac interferometer can control the light output at the same (different) side of the non-polarizing beam splitter (NPBS) where it is sent so that we can realize the quantum direct (common) cause. Observables $\hat{O}_A$ and $\hat{O}_B$ are measured to detect the causal structures. Notation for optical elements: polarizer (Pol.), liquid crystal retarder (LCR).}
	\label{fig:experiment}
\end{figure*}
{\em Experiment--}We implement a quantum optical experiment shown in Fig. \ref{fig:experiment} to verify the theoretical results. We encode the qubit in the polarization degree of freedom of single photon ($\ket{0}$ and $\ket{1}$ correspond to the horizontal $\ket{H}$ and vertical $\ket{V}$ polarization). The quantum states having fidelity around 96.4\% with the maximally entangled states $\ket{\Psi^-}=(\ket{HV}-\ket{VH})/\sqrt{2}$ are generated by using the spontaneous parametrical down-conversion (SPDC) (Fig.~\ref{fig:experiment} (a)). The waveplates controlled by a quantum random number generator (QRNG) are utilized to prepare different quantum states for further investigation.

To realize different causal structures between two quantum systems $A$ and $B$ in the experiment, we implement a displaced Sagnac interferometer shown in Fig. \ref{fig:experiment}(b). In this way, if the light measured with $\hat{O}_A$ outputs at the same (different) side of the non-polarizing beam splitter where it is sent, we can realize the quantum direct (common) cause. The mixing probability $p$ is set if the trial ratio of the direct cause to the whole experiment trials is $p$. The quantum direct cause imposed by different unitary operators $\hat{U}$ and their probabilistic mixtures can be achieved by the waveplates controlled by another QRNG. The observables $\hat{O}_A$ and $\hat{O}_B$ are measured at each systems and the coincidence is recorded by the two single photon counting modules (SPCM). (see experimental details in Ref. \cite{Supplementary_material}).

\begin{figure*}[t]
	 \centering
   \includegraphics[width=1\textwidth]{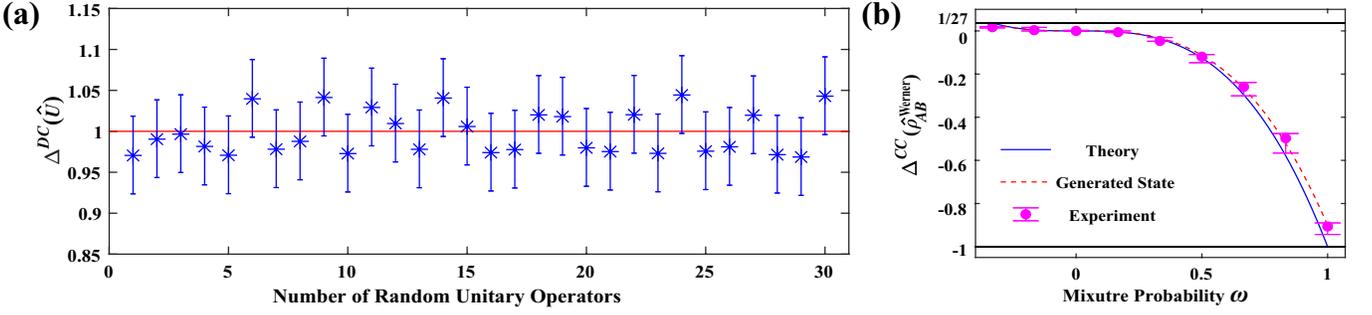}
   \caption[width=1\textwidth]{Experimental results of $\Delta^{DC}(\hat{U})$ and $\Delta^{CC}(\hat{\rho}_{AB})$. (a) Values of $\Delta^{DC}(\hat{U})$. Red line represents the theoretical value $\Delta^{DC}(\hat{U})\equiv 1$. Blue asterisks are the experimentally measured values. (b) The values of  $\Delta^{CC}(\hat{\rho}_{AB})$.
	 The black lines denote the boundary values of $\Delta^{CC}(\hat{\rho}_{AB})$. The blue-solid line is the theoretical $\Delta^{CC}(\hat{\rho}_{AB})$ for ideal Werner-like states. The theoretical predictions for the state generated in the experiment are denoted as the red-dashed line. The magenta dots are the experimentally measured values.}
	\label{fig:proveone}
\end{figure*}

{\em Results--}The experimentally measured causal determinant for quantum direct cause imposed by different unitary operators are shown in Fig. \ref{fig:proveone}(a) with blue asterisks. The error bars originate from the statistical fluctuations. The unitary operators are generated by utilizing the method of Haar measure \cite{ozols2009generate} to ensure they are uniformly distributed. The results confirm that $\Delta^{DC}(\hat{U})\equiv 1$ within the statistical fluctuation.

To demonstrate the results of the quantum common cause, we calculate the causal determinant for the well-studied Werner-like states \cite{PhysRevA.40.4277,PhysRevLett.84.4236,PhysRevLett.92.177901} $\hat{\rho}_{AB}^{\text{Werner}}=(1-\omega)(\hat{I}_A\otimes\hat{I}_B)/4+\omega\ketbra{\Psi^-}{\Psi^-}$, where
$\omega\in[-1/3,1]$ and $\hat{I}$ is the identity operator.
As the mixing probability $\omega$ changes, it is obvious that the causal determinant of quantum common cause described by Werner-like states can reach the whole range $\Delta^{CC}(\hat{\rho}_{AB}^{\text{Werner}})\in[-1,1/27]$ denoted by blue-solid line in Fig. \ref{fig:proveone}(b). In the experiment, the Werner-like states are generated by probabilistically mixing the four Bell states with the help of the QRNG.  After the measurement, the experimental causal determinants for Werner-like states with different mixing probability $\omega$ are obtained (magenta dots with error bars), which deviate a little from the theoretical predictions due to the imperfections of the states we generate.
We calculate the causal determinants for the states generated in the experiment (red-dashed line), which coincide well with our experimental results.

When the quantum direct cause is imposed by a more general quantum channel $\mathcal{E}$, the ranges of the causal determinant depend on the number of unitary operators $N^{DC}$. If $N^{DC}=1$, we have $\Delta^{DC}\equiv 1$. The quantum channel $\mathcal{E}$ is reduced to the unitary evolution. However, if $N^{DC}>1$, the causal determinant are no longer a constant. For example, if $N^{DC}=2$, we have $\Delta^{DC}_{N^{DC}=2}\in[0,1]$. For the cases $ N^{DC}\geq 3$,
we obtain the range $\Delta^{DC}_{N^{DC}\geq3}\in[-1/27,1]$. Apparently, there exists overlap in the ranges of the causal determinant between quantum direct cause in the case of $N^{DC}>1$ and the quantum common cause so that these two causal mechanisms cannot be perfectly distinguished through the ranges of the causal determinant. Even though, the quantum direct cause can still be confirmed when the measured causal determinant lies in the range $(1/27,1]$. Besides, the range of the $N^{DC}$ can also be partially inferred according to the causal determinant. If $\Delta^{DC}\in[0,1)$, we have $N^{DC}\geq 2$.
If the causal determinant $\Delta^{DC}\in[-1/27,0)$, we know for sure that $N^{DC}\geq 3$.

\begin{figure}[t]
	 \centering
   \includegraphics[width=0.48\textwidth]{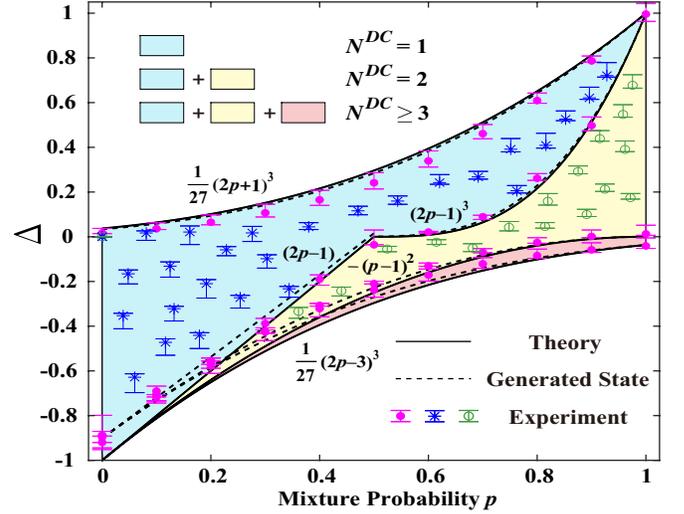}
   \caption[width=1\textwidth]{The values of $\Delta$ for general quantum causal structures. When $N^{DC}=1$, $\Delta$ should lie in the cyan region. If $N^{DC}=2$, the values are extended into the yellow region. The red region should also be included if $N^{DC}\geq 3$. The ideal boundaries for each regions are depicted as the black-solid lines. The boundaries for the states generated in the experiment are depicted as the red-dashed lines. The experimental results for boundaries are shown as the magenta dots. The blue asterisks denote the measured causal determinants for $N^{DC}=1$. For $N^{DC}=2$, the results lying in the yellow region denoted as green circles.}
	\label{fig:results}
\end{figure}

For more general quantum causal structures with probabilistic mixture of both common and direct causes, the range of the causal determinant $\Delta$ changes with the mixing probability $p$. In the cases of $p=0$ and $p=1$, we obtain the pure quantum common and direct cause, respectively. Otherwise, for $p\in(0,1)$, the causal determinant is related to the number of unitary operators $N^{DC}$ shown in Fig. \ref{fig:results}. If $N^{DC}=1$, the causal determinant is limited in the cyan region. If $N^{DC}=2$, the range of the causal determinant is extended to the yellow region. The red region can also be included if $N^{DC}\geq 3$. Clearly, different cases share a common upper bound. All the quantum causal structures with $N^{DC}\geq 3$ also share a common lower bounds. The values of the boundaries for each region can be obtained by certain quantum states and channels \cite{Supplementary_material}.

In the experiment, we firstly measure the causal determinant of the boundaries for each regions shown as magenta dots with error bars in Fig.~\ref{fig:results}. The deviation from the ideal bound (black-solid lines) originates from the imperfections of the states we generate. Again, we calculate the values of the causal determinant for the states generated in the experiment (black-dashed lines). The experimental results now agree well with the theoretical predictions. The regions surrounded by the boundaries can be filled by appropriately mixing the quantum direct and common causes, which are demonstrated in our experiment. When we mix the quantum direct cause in the case of $N^{DC}=1$ with the quantum common cause imposed by different quantum states, the measured causal determinant (blue asterisks with error bars) fill the cyan region. When $N^{DC}=2$, the causal determinants are extended to the yellow region shown as the green circles with error bars. Here, we only show the results in the yellow region for explicit.

The experimental results demonstrate the validity to infer the quantum causal structures with the causal determinant. Basically, we can witness both quantum direct and common cause using the causal determinant. If we obtain $\Delta> 1/27$, there must be contribution from the direct cause between two quantum systems. Besides, the larger value of $\Delta$ implies more contribution from the direct cause. On the contrary, $\Delta< -1/27$ suggests the presence of the common cause. The lower value implies more contribution from the common cause. Furthermore, for fixed causal determinant we obtained, we can predict the ranges of the mixing probability $p$ and the number of the unitary operators $N^{DC}$ \cite{Supplementary_material}.

{\em Discussion and Outlook--}In this work, we introduce a quantity
named `Causal Determinant', which enhances the ability to infer the causal structures between two correlated quantum systems. According to the causal determinant, the quantum direct cause imposed by a unitary operator can be perfectly discriminated with the quantum common causes imposed by any joint quantum states in an efficient way. For the quantum direct cause imposed by a more general quantum channel, the unital channel, the causal determinant still have certain capability to distinguish it with the quantum common cause. Moreover, for general quantum causal structures in which the probabilistic mixture of quantum direct and common causes is allowed, the causal determinant can help us to witness both causal mechanisms and predict the ranges of the mixing probability. These results are confirmed by a quantum optical experiment.

Future extension of the studies on the causal inference with the use of the causal determinant may cover the situation of more complex causal structures. For example, the coherent mixture of the quantum direct cause and the quantum common cause, the quantum direct cause that is imposed by any general quantum channels. On the other hand, our results may be extended to infer the causal structures between quantum systems with higher dimension.
\begin{acknowledgments}
  This work was supported by the National Key Research and Development Program of China under (Grant Nos. 2017YFA0303703 and 2018YFA0306202), the National Natural Science Foundation of China (Grant Nos. 11690032, 61490711, 11474159, 11574145, 11875160,12175104), Fundamental Research Funds for the Central
  Universities (Grant No. 021314380197).
\end{acknowledgments}
%\bibliography{reference}

\end{document}